\documentclass{article}

\usepackage{arxiv}

\usepackage[utf8]{inputenc} 
\usepackage[T1]{fontenc}    
\usepackage{url}            
\usepackage{booktabs}       
\usepackage{amsfonts}       
\usepackage{nicefrac}       
\usepackage{microtype}      
\usepackage{lipsum}		
\usepackage{graphicx}
\usepackage{natbib}
\usepackage{doi}

\usepackage{xurl}
\usepackage{hyperref}
\hypersetup{
colorlinks=true,
urlcolor=blue,
linkcolor=blue,
citecolor=blue
}
\usepackage{xcolor}
\definecolor{newcolor}{rgb}{.8,.349,.1}

\title{Siberian radioheliograph image classification using ensemble of CLIP, EfficientNet and CatBoost models}

\author{ \href{https://orcid.org/0000-0003-3993-1083}{\includegraphics[scale=0.09]{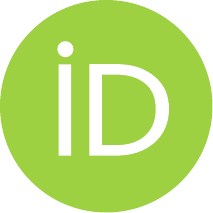}\hspace{1mm}Yaroslav~Egorov} \\
	Institute of Solar-Terrestrial Physics SB RAS\\	
	664033, Irkutsk, Russia \\
	\texttt{egorov@iszf.irk.ru} \\
}

\renewcommand{\shorttitle}{SRH image classification using ensemble model}

\hypersetup{
pdftitle={Siberian radioheliograph image classification using ensemble of CLIP, EfficientNet and CatBoost models},
pdfsubject={q-bio.NC, q-bio.QM},
pdfauthor={Yaroslav Egorov},
pdfkeywords={Sun, Radio emission, Radio telescopes, Siberian radioheliograph, Transfer learning, EfficientNet, CLIP, CatBoost, Ensemble model},
}

\begin{document}
\maketitle

\begin{abstract}
The Siberian Radioheliograph (SRH) is a ground-based radio interferometer in Irkutsk, Russia, designed for high-resolution solar observations in the microwave range. It can observe dynamic solar events with spatial resolutions of 7–30 arcseconds and temporal resolution up to 0.1 seconds.

Generating solar radio images from the Siberian Radioheliograph (SRH) is a multi-step calibration process that corrects instrumental and atmospheric distortions, using redundancy-based calibration with both adjacent and non-adjacent antenna pairs to address phase and amplitude errors in visibility data. The CLEAN algorithm is then applied to deconvolve the point spread function, reduce sidelobes, and enhance the visibility of solar features, resulting in higher quality and more reliable images.

While the calibration process generally improves image quality, it can sometimes result in noisy or spatially shifted images that are not suitable for scientific use. We developed a deep learning approach for automatic image quality classification. The training dataset was prepared using a zero-shot CLIP model and further validated manually. We evaluated four different models: a fine-tuned EfficientNet, two CatBoost variants using embeddings from CLIP and EfficientNet, and an Ensemble model that combined predictions from all three individual models. The Ensemble model achieved the best performance.

The SRH daily image classification service has been created and is available online at \url{https://forecasting.iszf.irk.ru/srh} along with an API offering IDL and Python examples. Integration of Ensemble model into SRH image generating and calibration workflow can improve image reliability and reduces low-quality entries in SRH data catalog, enhancing solar research outcomes.
\end{abstract}


\section{Introduction}
\label{Introduction}

Solar activity poses significant risks to Earth’s technology and environment, including geomagnetic storms that can disrupt satellites and power grids, radiation hazards that endanger astronauts and spacecraft, and communication failures that degrade the reliability of radio systems and GPS accuracy \citep{Bala2002,Kutiev2013,Knipp2016,Marque2018}. Increased atmospheric drag can alter the trajectory of satellites, increasing the risk of collisions, and penetrating particles can damage spacecraft electronics. Hard electromagnetic emission and high-energy particles produced by solar flares can directly affect the ionosphere \citep{Afraimovich2001, Tsurutani2009,Yasyukevich2018}. In addition, cosmic rays become a greater threat to high-altitude aviation during periods of reduced solar activity, and geomagnetically induced currents can lead to power grid failures and corrosion of pipelines \citep{Barlow1849,Campbell1980,Viljanen2006}. These consequences emphasize the need to investigate the physical processes of solar activity and its major phenomena, such as solar flares, coronal mass ejections (CMEs), high-energy solar particles and interplanetary effects.

Solar atmosphere heating mechanisms, including both gradual and impulsive processes, can be studied more effectively with a multi-wavelength approach. It helps in establishing connections between radio emissions and heating processes observed in X-ray and UV \citep{Altyntsev2020multiwave}.

Integrated data enables refined models of solar activity that can accurately predict solar flares, coronal mass ejections, and their effects on space weather \citep{Grechnev2014}. Understanding the interplay of emissions across different wavelengths enhances forecasting capabilities, which is crucial for mitigating the affects of space weather on technology and infrastructure on Earth.

The Siberian Radioheliograph (SRH) is a ground-based radio interferometer developed by the Institute of Solar-Terrestrial Physics SB RAS in Irkutsk, Russia. Designed for high-resolution, multi-frequency solar observations in the microwave range (3–24 GHz) with spatial resolution ranging from 7 to 30 arcseconds and temporal resolution up to 0.1 seconds \citep{Lesovoi2017siberian, Altyntsev2020multiwave}. 

The SRH can monitor indices of solar activity, such as the $F_{10.7}$ index, and detect narrow-band radio bursts. A number of modern empirical atmospheric models use the $F_{10.7}$ index as a proxy for solar extreme ultraviolet (EUV) flux \citep{picone, Bowman2008, Bruinsma2015}. Accurately measuring and predicting solar activity is crucial for understanding space weather phenomena, which significantly impact satellite operations, communication systems, and terrestrial climate dynamics. In recent years, various machine learning methods have been successfully applied to predict $F_{10.7}$ \citep{Zhang2022, STEVENSON2022, hao, EGOROV2025455}.

Accurate solar radio images from the Siberian Radioheliograph (SRH) are created through a multi-step calibration process that corrects for instrumental and atmospheric distortions \citep{Lesovoi2017siberian, Fedotova2019calibration}. This calibration addresses phase and amplitude errors in visibility data, which can otherwise reduce image quality and dynamic range. 

The SRH uses redundancy-based calibration, relying on multiple antenna pairs measuring the same spatial frequencies, eliminating the need for external calibrators. By including both adjacent and non-adjacent antenna pairs, this method enables precise corrections and more accurate imaging of extended sources like the quiet Sun \citep{Lesovoi2017siberian, Fedotova2019calibration, Globa2021calibration}. The CLEAN algorithm \citep{Hogbom1974} is then applied to further refine the images by deconvolving the point spread function from the observed brightness distribution, effectively reducing side lobes and enhancing the visibility of solar features. The cleaned images (CLEAN MAPS) are available at \url{https://ftp.rao.istp.ac.ru/SRH/SRH0306/cleanMaps}.

The development of an automated method for assessing image quality is essential, as it would enhance the self-calibration technique by providing validation of results during its iterative process. Furthermore, such a method could facilitate the filtering of data in catalog (\url{https://ftp.rao.istp.ac.ru/SRH/SRH0306/cleanMaps}), ensuring that only high-quality data are used in scientific analyses.

The implementation of deep learning in scientific research has revolutionized the analysis of images across various domains, including astronomy, medicine, and environmental science \citep{Yu2021,Ramos2023}. Deep learning models, such as convolutional neural networks (CNNs), have demonstrated exceptional performance in tasks like image classification, object detection, and segmentation because of their ability to automatically learn hierarchical features from raw data. For example, \cite{Illarionov2018} and \cite{Jarolim2021} used CNN to identify coronal hole boundaries. In radio astronomy, CNNs can be employed to distinguish between high-quality and low-quality images by learning patterns associated with noise levels, resolution, and artifacts. 

Last years, researchers increasingly adopt these techniques to address complex challenges in image-based studies, leveraging pre-trained architectures for transfer learning \citep{Armstrong2019,Siahkoohi2019}. Transfer learning is a machine learning technique where a model pre-trained on one task is reused as the starting point for another related task. It is particularly useful when the target dataset is small or lacks sufficient diversity to train a robust model from scratch \citep{Hosna2022}.

In this study, we propose an ensemble model that combines predictions from the CLIP \citep{clip}, EfficientNet \citep{tan2020efficientnetrethinkingmodelscaling}, and CatBoost \citep{catboost} models to classify the image quality of Siberian Radioheliograph data. By combining the complementary strengths of these models, our ensemble approach improves the classification accuracy and robustness, offering a more reliable and consistent image quality assessment than any single model. This work represents a new application of a multi-model ensemble in solar radio image analysis, making a valuable contribution to automated quality control in Siberian Radioheliograph data.

\section{Data and Method}
\label{data}

\subsection{SRH image data}

The SRH catalog (\url{https://ftp.rao.istp.ac.ru}) contains more than 100 000 radio images from August 11, 2021 to the present with a time cadence of about 2 min. Data is available in different microwave range from 3 to 24 GHz. Besides good quality images, it contains shifted and noisy images. In this paper, we will classify only "GOOD" images and "BAD" ones for 3 GHz data. Thus, shifted, fuzzy and noisy images fall into the "BAD" class.

SRH images retrieved from FITS files are preprocessed to enhance contrast, normalize intensity values, and ensure numerical stability for subsequent analysis. The adjustment procedure is implemented as follows:

\textbf{Data Loading}: FITS file data is read using the astropy.io.fits

\textbf{Finite Value Masking}: A mask is generated to identify valid pixel values, excluding NaNs or infinities that may result from observational artifacts or data corruption.

\textbf{Normalization}: The image data ($I$) is normalized by subtracting the mean ($\mu$) and dividing by the standard deviation ($\sigma$) of valid pixels, calculated while ignoring non-finite values:

\begin{equation}
    I_{N} = \frac{I - \mu}{\sigma}
\end{equation}

\textbf{Logarithmic Transformation}: To enhance low-intensity features and compress dynamic range, the normalized data undergoes a log-plus-one transformation:

\begin{equation}
    I_{adj}=\log(1+(I_{N} \cdot \min(I_{N})))
\end{equation}

\textbf{Saving images}: Enhanced data is saved in PNG format for subsequent machine learning tasks.

\begin{figure}[!ht]
\centering
    {\includegraphics[width=0.99\textwidth, trim=0.0cm 0cm 0cm 0cm]{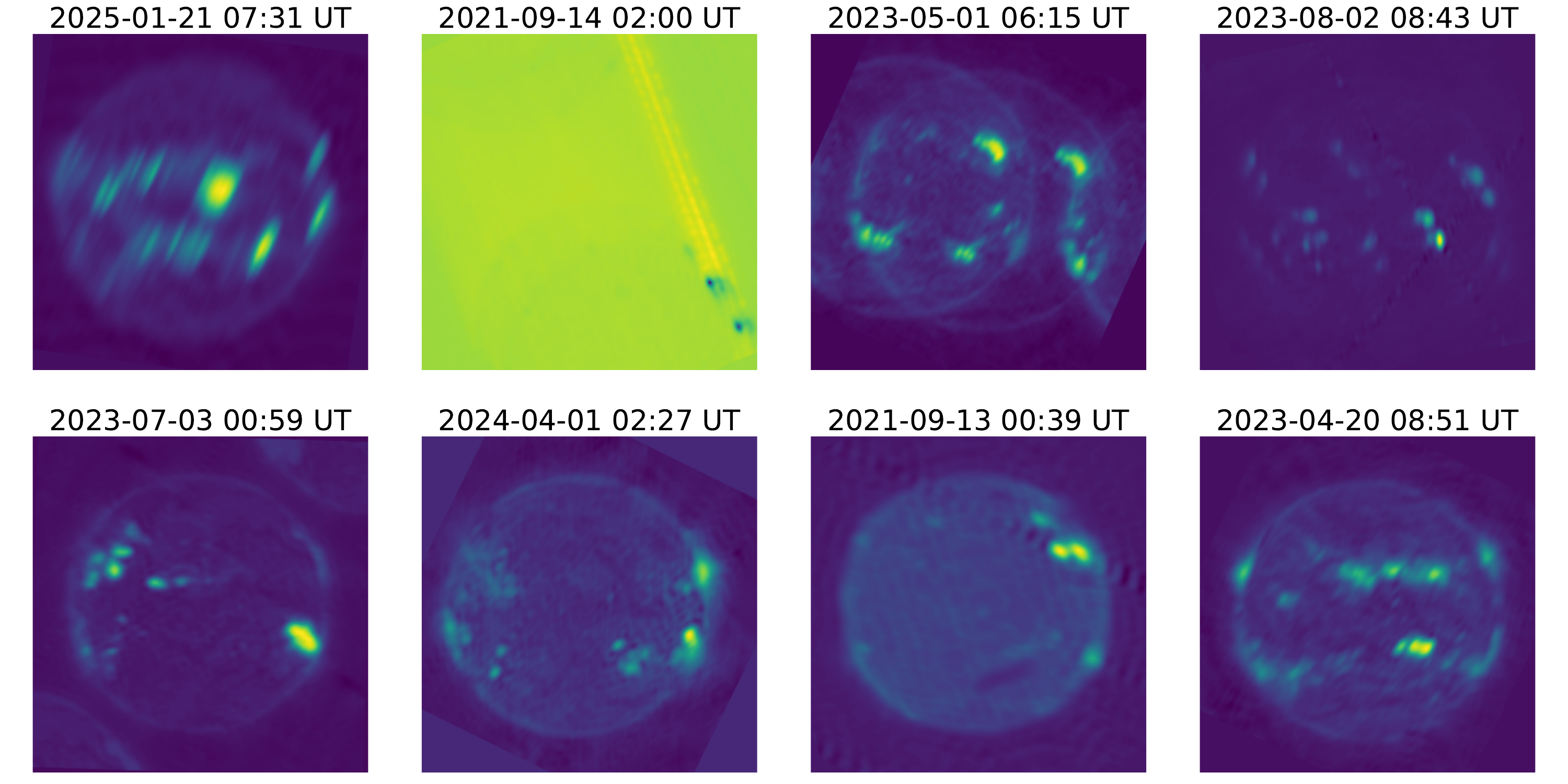}}
    \caption{The images from the SRH catalog represent typical cases used for classification, with "BAD" examples showing poor quality (top panel), and "GOOD" examples demonstrating high-quality (bottom panel)}
    \label{fig:examples}
\end{figure}

To fine-tune the EfficientNet, one should prepare the labeled training dataset with “GOOD” and “BAD” classes. The accuracy of a model increases with the size of a training dataset. Preparing a large dataset is time consuming.

\subsection{Preparing data for image classification with CLIP}

Zero-shot Contrastive Language-Image Pre-training (CLIP) model \citep{clip} is able to understand both images and text to automatically assign labels to data on their content. CLIP is a pre-trained model that learns to associate images with text descriptions by training on a large dataset of 400 million image-text pairs obtained from the web.

\begin{figure}[!ht]
\centering
    {\includegraphics[width=0.7\textwidth, trim=0.0cm 0cm 0cm 0cm]{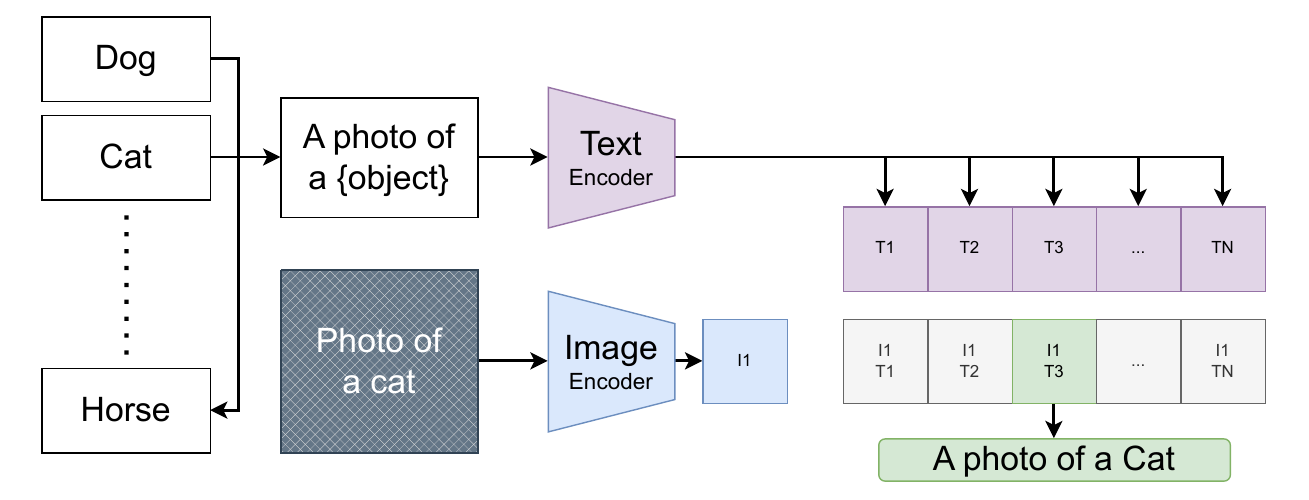}}
    \caption{A multimodal architecture of CLIP for image and text encoding. The model consists of separate encoders for images and text, mapping them into a shared embedding space. }
    \label{fig:clip}
\end{figure}

To label the data with CLIP, we provided two class labels as text cues (“photo of a circle” for the “GOOD” class and “photo of noise” for the “BAD” class) and used CLIP to compute similarity scores between each image and the text cues. This process eliminates the need for manual labeling, saving time and effort, especially for a large dataset like the SRH catalog. Figure \ref{fig:clip} shows a scheme of the CLIP model, which encodes both text and image in a shared embedding space. This innovative learning strategy allows CLIP to generalize to a wide range of tasks, achieving performance levels comparable to fine-tuned supervised models.

With CLIP we prepared 100 000 images from SRH catalog and then manually verified a dataset of 10000 images. It divided into three datasets: training (6000), validation (2000) and test (2000) with an equal number of images for each class. The test dataset will be used to evaluate the models we use. In the next step, the training and validation datasets will be used to fine tune the pre-trained EfficientNet model. 

\subsection{Fine-tuning EfficientNet}

Transfer learning is a technique that uses the knowledge gained from training a model on one task to improve its performance on another, but related task, rather than starting the training process from scratch. This approach is especially useful when limited data is available for the target task, as it allows the model to use pre-trained features learned from large datasets. We used the EfficientNet-B0 trained on the ImageNet dataset \citep{effnet} to SRH image classification task using transfer learning. 

Data preparation follows a standard process, with separate training, validation, and test sets organized by class. Images are resized to a uniform dimension, normalized using ImageNet statistics, and augmented using random cropping and flipping during training. The same resizing is applied consistently during validation and testing.

Training is performed in two stages: first, only the classifier is trained while the feature extractor layers are kept frozen. Then, the top layers of the network are unfrozen for fine-tuning at a reduced learning rate. The framework also supports the extraction of embeddings, which allows for analysis of the learned representations. These embeddings can be used for downstream tasks such as clustering, visualization, and, in this study, as input to a CatBoost model.


\subsection{CatBoost}

Both CLIP and EfficientNet encode images into embeddings, which are numerical representations of the semantic content of an image. These embeddings can be used in other models, such as CatBoost, to take advantage of different types of neural networks.

CatBoost is a gradient boosting framework developed by Yandex and optimized for tabular data \citep{catboost}. Once image embeddings—typically represented as fixed-length dense vectors—are extracted, they resemble tabular features. CatBoost excels at modeling such structured data, often outperforming other classical classifiers like Random Forests due to its advanced handling of decision trees, native support for GPU acceleration, and built-in cross-validation.

CatBoost includes strong regularization mechanisms such as ordered boosting, permutation-driven overfitting control, and automatic handling of categorical features. These help mitigate overfitting when training on relatively small sets of embeddings, which is especially valuable when working with limited labeled datasets in specialized domains like solar physics or astrophysics.

\subsection{Ensemble Model}
\label{sub:ensemble}

To combine predictions from multiple base models into a unified decision, we implement a lightweight feedforward neural network (FFNN) as an ensemble classifier. This model takes as input a 6-dimensional feature vector — representing class predictions from Fine-tuned EfficientNet, CatBoost + CLIP and CatBoost + EfficientNet  — and outputs a final binary classification decision ("Good" or "Bad").

\begin{figure}[!ht]
\centering
    {\includegraphics[width=0.59\textwidth, trim=0.0cm 0cm 0cm 0cm]{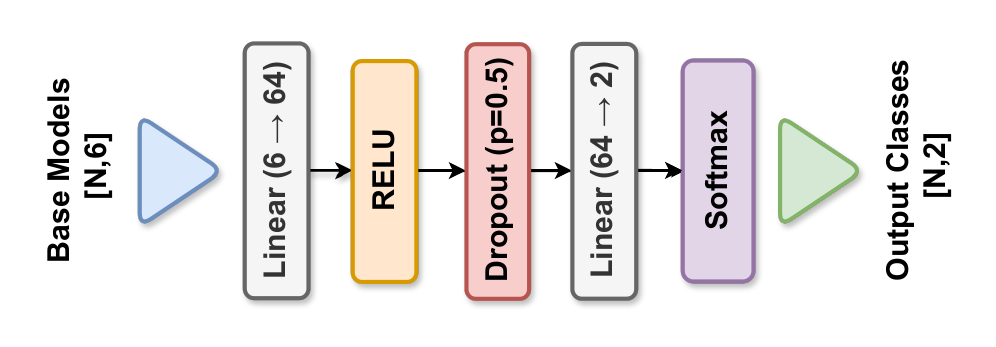}}
    \caption{ Architecture of the Ensemble Classifier. The model takes outputs from 3 base models as input and processes them through a two-layer feedforward neural network with ReLU activation and dropout regularization (dropout rate = 0.5). A final softmax layer produces class probabilities for two output categories: "Bad" and "Good".}
    \label{fig:ensemble}
\end{figure}

The architecture consists of two fully connected (linear) layers with ReLU activation and dropout regularization to prevent overfitting (Figure \ref{fig:ensemble}). This design enables the model to learn non-linear combinations of the input features while maintaining simplicity and computational efficiency.

\subsection{Workflow}
\label{sub:workflow}

\begin{figure}[!ht]
\centering
    {\includegraphics[width=0.89\textwidth, trim=0.0cm 0cm 0cm 0cm]{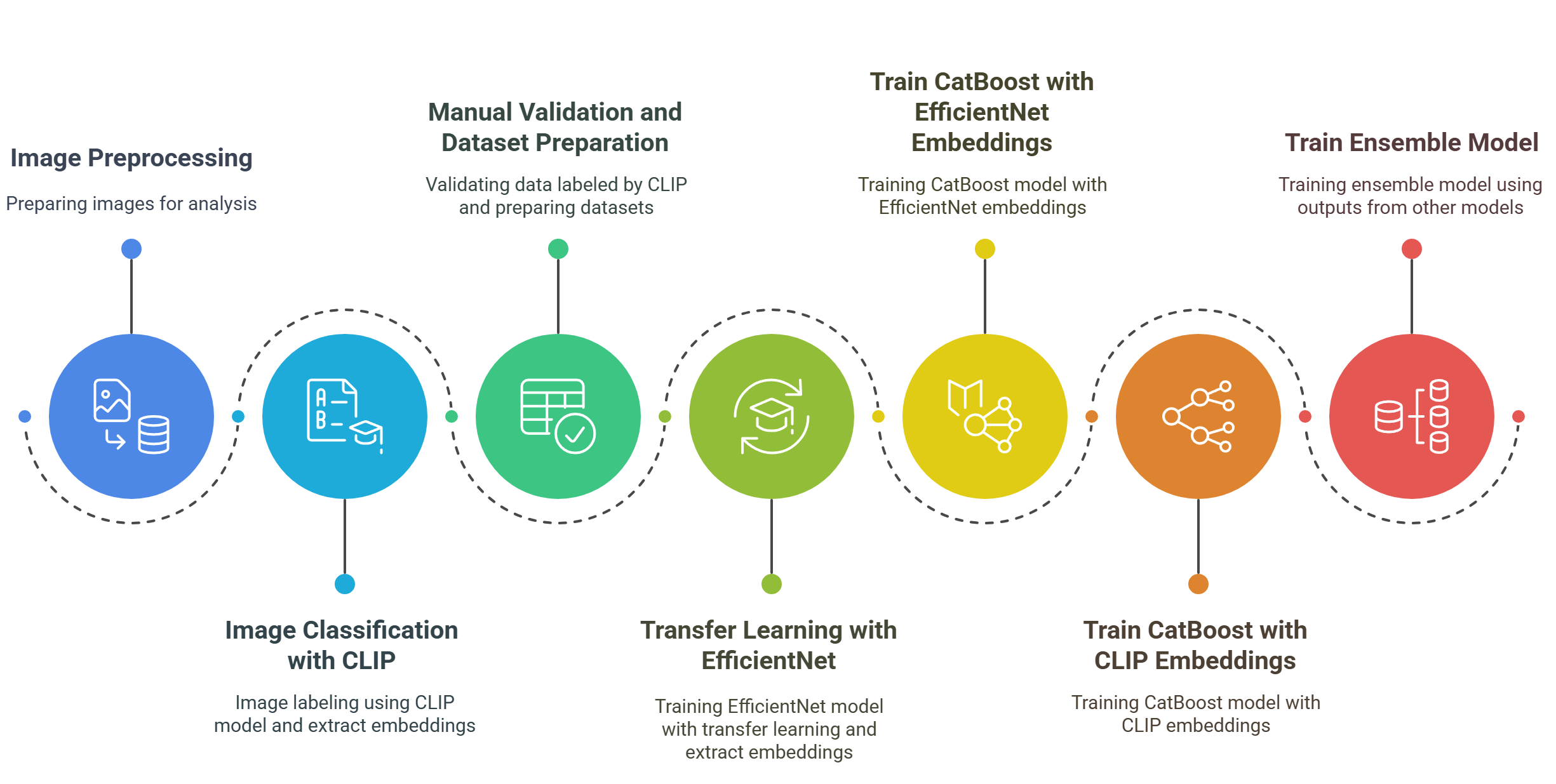}}
    \caption{Illustration for processing and classifying solar radio images using a combination of image preprocessing, CLIP-based classification, transfer learning with EfficientNet, CatBoost models, and ensemble techniques. The workflow includes data preparation, manual validation, model training, embedding extraction, and ensemble classification to improve performance.}
    \label{fig:workflow}
\end{figure}

The workflow consists of six main stages (Figure \ref{fig:workflow}). The process begins with Image Preprocessing , where raw images are resized, normalized, and augmented to ensure consistency and robustness in the data.

Next, the CLIP Model is used for image classification. This stage involves annotating the images and extracting semantic embeddings—high-level feature representations that capture meaningful visual information for use in downstream tasks.

Following this, the dataset generated by CLIP undergoes Manual Validation , ensuring accuracy and reliability of the labels. The dataset is then split into training, validation, and test sets for model development and evaluation.

In the next stage, the EfficientNet model is trained using transfer learning techniques and extracts embeddings to use it in later steps.

The embeddings—both from CLIP and EfficientNet—are then used to train two separate CatBoost models : one utilizing CLIP embeddings and the other using EfficientNet embeddings. These gradient-boosted models learn from the high-dimensional feature representations to improve classification performance.

Finally, all previous models contribute their predictions to an ensemble model, which combines them to improve overall performance by leveraging the strengths of the models used. All code, trained models, and evaluation metrics are available in the GitHub repository \url{https://github.com/EgorovYaroslav/SRH.git}.

\subsection{Evaluation Metrics} 

A confusion matrix is a fundamental tool in machine learning and statistics used to evaluate the performance of classification models by comparing actual and predicted class labels. For binary classification, it consists of four outcomes: true positives (TP) , true negatives (TN) , false positives (FP) , and false negatives (FN). It can be extended to multi-class problems by representing each class as a row and column in the matrix -- it reveals patterns of correct predictions and misclassifications across all classes.

From the confusion matrix, key performance metrics are derived:

\textbf{Accuracy}: Measures overall correctness of predictions:

\begin{equation}
    Accuracy = \frac{TP+TN}{TP+FP+TN+FN}    
\end{equation}

\textbf{Precision}: Reflects the proportion of true positives among all positive predictions:

\begin{equation}
    Precision = \frac{TP}{FP+TP}
\end{equation}

\textbf{Recall}: Indicates the model’s ability to identify all positive instances:

\begin{equation}
    Recall = \frac{TP}{TP+FN}
\end{equation}

\textbf{F1-score}: A harmonic mean of precision and recall, particularly useful for imbalanced datasets:

\begin{equation}
    F1\_score = \frac{2TP}{2TP+FP+FN}
\end{equation}

These metrics provide a detailed understanding of model behavior, especially in contexts where class imbalance or error types have different consequences. In the Results section, we compared the performance of the models using the confusion matrix and the evaluation metrics described.

\section{Results}
\label{results}

\subsection{Comparison of classification performance}

To identify the optimal model for SRH image classification, we evaluate the performance of each model using various metrics on the test dataset. Next models has been compared:
\begin{itemize}
    \item Fine-tuned EfficientNet B0
    \item Catboost with embeddings encoded by CLIP
    \item Catboost with embeddings encoded by EfficientNet B0.
    \item Ensemble model with architecture described in Section \ref{sub:ensemble}
\end{itemize}
 
\begin{figure}[!ht]
\centering
    {\includegraphics[width=0.99\textwidth, trim=0.0cm 0cm 0cm 0cm]{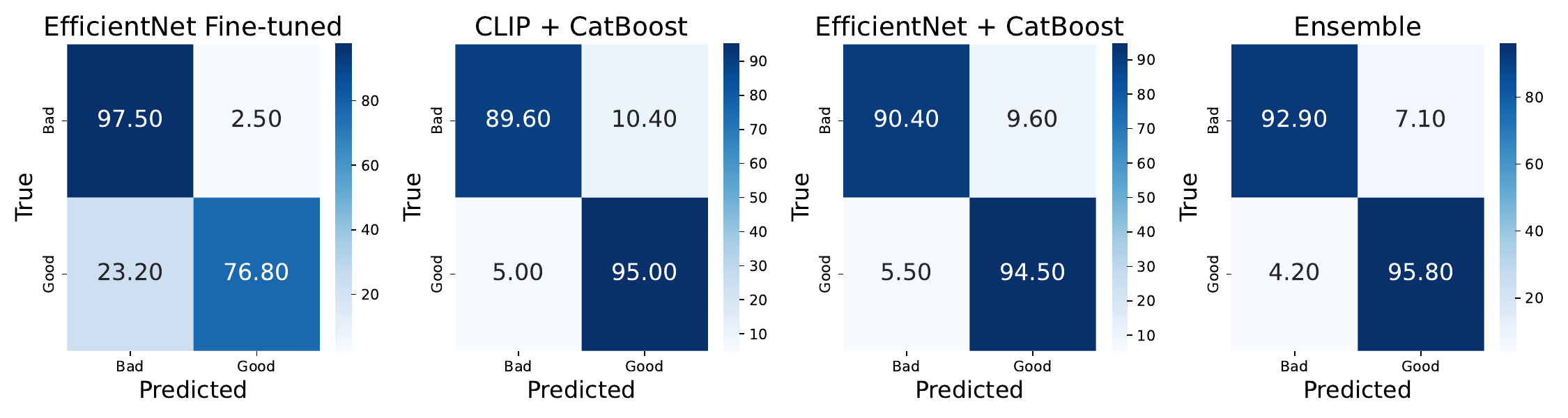}}
    \caption{Performance comparison of different models on classification tasks. The results show the accuracy (\%) and error rates (\%) for four different approaches: (1) CLIP, (2) Fine-tuned EfficientNet, (3) CLIP embeddings + CatBoost, and (4) EfficientNet embeddings + CatBoost. The models are evaluated on two classes: "Good" and "Bad". Each model's prediction accuracy and confusion matrix percentages are displayed, illustrating the superior performance of ensemble methods in classification accuracy}
    \label{fig:img3}
\end{figure}

Figure \ref{fig:img3} presents comparison of confusion matrix across different model architectures. We found that the hybrid approach provides high classification performance with accuracy 92\%, and ensemble methods further improve this metrics up to 95\%.

\begin{figure}[!ht]
\centering
    {\includegraphics[width=0.99\textwidth, trim=0.0cm 0cm 0cm 0cm]{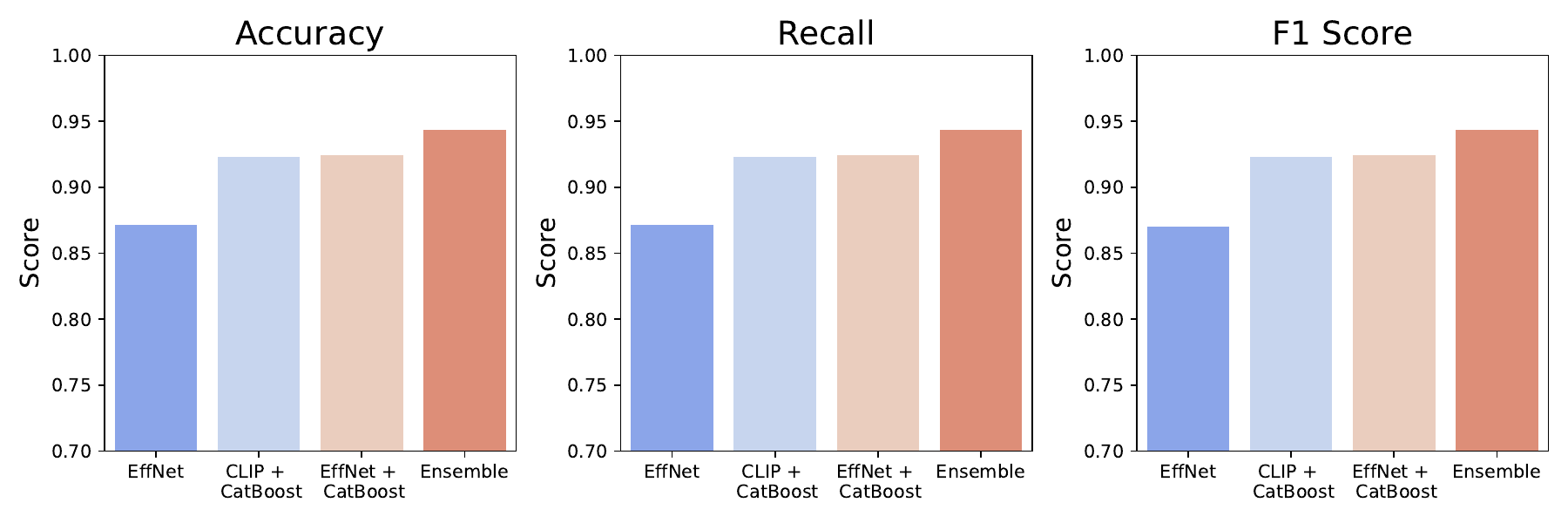}}
    \caption{Comparison of classification performance metrics across different model architectures. The plot shows Accuracy , Recall , and F1 Score for four approaches: (1) CLIP, (2) EfficientNet (EffNet), (3) CLIP combined with CatBoost, and (4) EfficientNet combined with CatBoost. All metrics are displayed on a scale from 0.70 to 1.00, highlighting the improved performance of ensemble methods in multi-class classification tasks.}
    \label{fig:img4}
\end{figure}

Figure \ref{fig:img4} shows a comparison across three key evaluation metrics: precision, recall, and F1 score for all models. The Ensemble model achieves the highest evaluation metrics, followed by the EfficientNet + CatBoost combination. CLIP + CatBoost performs slightly worse, while the standalone EfficientNet shows the lowest metrics.

\subsection{Service for regular SRH image classification }

\begin{figure}[!ht]
\centering
    {\includegraphics[width=0.79\textwidth, trim=0.0cm 0cm 0cm 0cm]{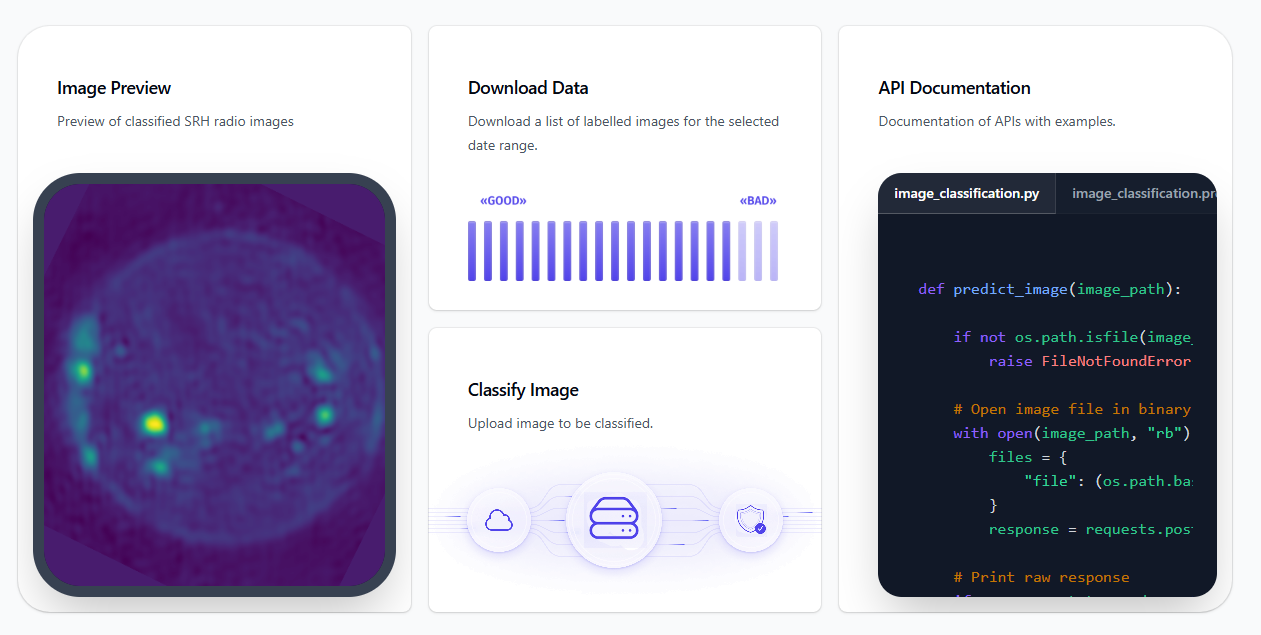}}
    \caption{Web interface of the SRH (Siberian Radioheliograph) radio image classification service, showcasing its key functionalities. The left panel displays an Image Preview, where users can view classified SRH radio images with labeled regions indicating "GOOD" or "BAD" quality. The middle panel highlights the Download Data feature, allowing users to download lists of labeled images for selected date ranges. The right panel presents the API Documentation , providing code examples in Python for integrating the service programmatically via its RESTful API. This interface facilitates efficient data exploration, classification, and automation for solar radio imaging analysis.}
    \label{fig:webservice}
\end{figure}

We selected the Ensemble model for daily classification of new SRH data and developed a web service available at \url{https://forecasting.iszf.irk.ru/srh} (Figure \ref{fig:webservice}). Users can upload FITS files for classification, explore results via a web interface, filter data by date or quality, and export metadata in CSV format. The service also includes a manual validation tool, allowing experts to review and refine classifications for improved accuracy. Additionally, it offers a RESTful API with example clients in Python and IDL to support integration into scientific workflows.

One potential application of the service is the integration of image classification via the API into the routine generation of CLEAN maps, which are available on the SRH website (\url{https://ftp.rao.istp.ac.ru}). Given the service’s high-speed classification capability, this integration may contribute to an improvement in the quality and reliability of the CLEAN maps.

\section{Conclusion}
\label{conclusion}

The Siberian Radioheliograph (SRH), developed by the Institute of Solar-Terrestrial Physics SB RAS in Irkutsk, Russia, is a ground-based radio interferometer designed for high-resolution, multi-frequency solar observations in the microwave range—offering spatial resolutions of 7–30 arcseconds and temporal resolution up to 0.1 seconds — to enable detailed studies of solar activity, monitor solar indices, detect radio bursts, and improve forecasting of solar flares and coronal mass ejections.

Generating radio images from source data is a complex process in radio astronomy that involves collecting visibility data through interferometric observations, calibrating the data to correct for instrumental and environmental effects, transforming it into an initial "dirty" image using a Fourier transform, and then improving the image quality with the CLEAN algorithm and self-calibration.

The development of an automated image quality assessment method is essential for improving self-calibration by validating results during its iterative process and ensuring only high-quality data are used in scientific analyses through efficient catalog filtering.

In this study, we propose a deep learning approach to classify image quality of SRH data. To prepare the training dataset we used zero-shot CLIP model that able to classify images using text description. The final dataset was validated manually and divided into train, validation and test datasets. 

We compared the image classification performance for 4 models, including Fine-tuned EfficientNet B0, two variants of Catboost model trained on image embeddings obtained with CLIP (CLIP + Catboost) and EfficientNet (EfficientNet + Catboost) and Ensemble model. The last model is a simple feedforward neural network that combines the predictions of three other models, achieving the best performance compared to any individual model.

The Ensemble model was chosen for regular daily SRH data classification and available online at \url{https://forecasting.iszf.irk.ru/srh}. Also this website contains API with IDL and Python examples that can classify SRH raw fits file. It can be used inside SRH image generating routine to control and correct the quality of resulting images. It should reduce the low-quality data in SRH catalog that can be used for scientific research.

\section*{Acknowledgments}
The author is grateful to the Siberian Radioheliograph for providing data. The work was supported by the Ministry of Science and Higher Education of the Russian Federation.

\bibliographystyle{unsrtnat}
\bibliography{refs}

\begin{thebibliography}{34}
\providecommand{\natexlab}[1]{#1}
\providecommand{\url}[1]{\texttt{#1}}
\expandafter\ifx\csname urlstyle\endcsname\relax
  \providecommand{\doi}[1]{doi: #1}\else
  \providecommand{\doi}{doi: \begingroup \urlstyle{rm}\Url}\fi

\bibitem[Bala et~al.(2002)Bala, Lanzerotti, Gary, and Thomson]{Bala2002}
B.~Bala, L.~J. Lanzerotti, D.~E. Gary, and D.~J. Thomson.
\newblock Noise in wireless systems produced by solar radio bursts.
\newblock \emph{Radio Science}, 37\penalty0 (2):\penalty0 2--1--2--7, 2002.
\newblock \doi{https://doi.org/10.1029/2001RS002481}.
\newblock URL
  \url{https://agupubs.onlinelibrary.wiley.com/doi/abs/10.1029/2001RS002481}.

\bibitem[{Kutiev} et~al.(2013){Kutiev}, {Tsagouri, Ioanna}, {Perrone,
  Loredana}, {Pancheva, Dora}, {Mukhtarov, Plamen}, {Mikhailov, Andrei},
  {Lastovicka, Jan}, {Jakowski, Norbert}, {Buresova, Dalia}, {Blanch,
  Estefania}, {Andonov, Borislav}, {Altadill, David}, {Magdaleno, Sergio},
  {Parisi, Mario}, and {Miquel Torta, Joan}]{Kutiev2013}
{Kutiev}, {Tsagouri, Ioanna}, {Perrone, Loredana}, {Pancheva, Dora},
  {Mukhtarov, Plamen}, {Mikhailov, Andrei}, {Lastovicka, Jan}, {Jakowski,
  Norbert}, {Buresova, Dalia}, {Blanch, Estefania}, {Andonov, Borislav},
  {Altadill, David}, {Magdaleno, Sergio}, {Parisi, Mario}, and {Miquel Torta,
  Joan}.
\newblock Solar activity impact on the earth’s upper atmosphere.
\newblock \emph{J. Space Weather Space Clim.}, 3:\penalty0 A06, 2013.
\newblock \doi{10.1051/swsc/2013028}.
\newblock URL \url{https://doi.org/10.1051/swsc/2013028}.

\bibitem[Knipp et~al.(2016)Knipp, Ramsay, Beard, Boright, Cade, Hewins,
  McFadden, Denig, Kilcommons, Shea, and Smart]{Knipp2016}
D.~J. Knipp, A.~C. Ramsay, E.~D. Beard, A.~L. Boright, W.~B. Cade, I.~M.
  Hewins, R.~H. McFadden, W.~F. Denig, L.~M. Kilcommons, M.~A. Shea, and D.~F.
  Smart.
\newblock The may 1967 great storm and radio disruption event: Extreme space
  weather and extraordinary responses.
\newblock \emph{Space Weather}, 14\penalty0 (9):\penalty0 614--633, 2016.
\newblock \doi{https://doi.org/10.1002/2016SW001423}.
\newblock URL
  \url{https://agupubs.onlinelibrary.wiley.com/doi/abs/10.1002/2016SW001423}.

\bibitem[{Marqué, Christophe} et~al.(2018){Marqué, Christophe}, {Klein,
  Karl-Ludwig}, {Monstein, Christian}, {Opgenoorth, Hermann}, {Pulkkinen,
  Antti}, {Buchert, Stephan}, {Krucker, Säm}, {Van Hoof, Rudiger}, and
  {Thulesen, Peter}]{Marque2018}
{Marqué, Christophe}, {Klein, Karl-Ludwig}, {Monstein, Christian},
  {Opgenoorth, Hermann}, {Pulkkinen, Antti}, {Buchert, Stephan}, {Krucker,
  Säm}, {Van Hoof, Rudiger}, and {Thulesen, Peter}.
\newblock Solar radio emission as a disturbance of aeronautical
  radionavigation.
\newblock \emph{J. Space Weather Space Clim.}, 8:\penalty0 A42, 2018.
\newblock \doi{10.1051/swsc/2018029}.
\newblock URL \url{https://doi.org/10.1051/swsc/2018029}.

\bibitem[Afraimovich et~al.(2001)Afraimovich, Altyntsev, Kosogorov, Larina, and
  Leonovich]{Afraimovich2001}
E.L. Afraimovich, A.T. Altyntsev, E.A. Kosogorov, N.S. Larina, and L.A.
  Leonovich.
\newblock Ionospheric effects of the solar flares of september 23, 1998 and
  july 29, 1999 as deduced from global gps network data.
\newblock \emph{Journal of Atmospheric and Solar-Terrestrial Physics},
  63\penalty0 (17):\penalty0 1841--1849, 2001.
\newblock ISSN 1364-6826.
\newblock \doi{https://doi.org/10.1016/S1364-6826(01)00060-8}.
\newblock URL
  \url{https://www.sciencedirect.com/science/article/pii/S1364682601000608}.

\bibitem[Tsurutani et~al.(2009)Tsurutani, Verkhoglyadova, Mannucci, Lakhina,
  Li, and Zank]{Tsurutani2009}
B.~T. Tsurutani, O.~P. Verkhoglyadova, A.~J. Mannucci, G.~S. Lakhina, G.~Li,
  and G.~P. Zank.
\newblock A brief review of “solar flare effects” on the ionosphere.
\newblock \emph{Radio Science}, 44\penalty0 (1), 2009.
\newblock \doi{https://doi.org/10.1029/2008RS004029}.
\newblock URL
  \url{https://agupubs.onlinelibrary.wiley.com/doi/abs/10.1029/2008RS004029}.

\bibitem[Yasyukevich et~al.(2018)Yasyukevich, Astafyeva, Padokhin, Ivanova,
  Syrovatskii, and Podlesnyi]{Yasyukevich2018}
Y.~Yasyukevich, E.~Astafyeva, A.~Padokhin, V.~Ivanova, S.~Syrovatskii, and
  A.~Podlesnyi.
\newblock The 6 september 2017 x-class solar flares and their impacts on the
  ionosphere, gnss, and hf radio wave propagation.
\newblock \emph{Space Weather}, 16\penalty0 (8):\penalty0 1013--1027, 2018.
\newblock \doi{https://doi.org/10.1029/2018SW001932}.
\newblock URL
  \url{https://agupubs.onlinelibrary.wiley.com/doi/abs/10.1029/2018SW001932}.

\bibitem[Barlow et~al.(1849)Barlow, Barlow, and Culley]{Barlow1849}
William~Henry Barlow, Peter Barlow, and R.~S. Culley.
\newblock Vi. on the spontaneous electrical currents observed in the wires of
  the electric telegraph.
\newblock \emph{Philosophical Transactions of the Royal Society of London},
  139:\penalty0 61--72, 1849.
\newblock \doi{10.1098/rstl.1849.0006}.
\newblock URL
  \url{https://royalsocietypublishing.org/doi/abs/10.1098/rstl.1849.0006}.

\bibitem[Campbell(1980)]{Campbell1980}
Wallace~H. Campbell.
\newblock Observation of electric currents in the alaska oil pipeline resulting
  from auroral electrojet current sources.
\newblock \emph{Geophysical Journal International}, 61\penalty0 (2):\penalty0
  437--449, 05 1980.
\newblock ISSN 0956-540X.
\newblock \doi{10.1111/j.1365-246X.1980.tb04325.x}.
\newblock URL \url{https://doi.org/10.1111/j.1365-246X.1980.tb04325.x}.

\bibitem[Viljanen et~al.(2006)Viljanen, Pulkkinen, Pirjola, Pajunpää, Posio,
  and Koistinen]{Viljanen2006}
A.~Viljanen, A.~Pulkkinen, R.~Pirjola, K.~Pajunpää, P.~Posio, and
  A.~Koistinen.
\newblock Recordings of geomagnetically induced currents and a nowcasting
  service of the finnish natural gas pipeline system.
\newblock \emph{Space Weather}, 4\penalty0 (10), 2006.
\newblock \doi{https://doi.org/10.1029/2006SW000234}.
\newblock URL
  \url{https://agupubs.onlinelibrary.wiley.com/doi/abs/10.1029/2006SW000234}.

\bibitem[Altyntsev et~al.(2020)Altyntsev, Lesovoi, Globa, Gubin, Kochanov,
  Grechnev, Ivanov, Kobets, Meshalkina, Muratov, Prosovetsky, Myshyakov,
  Uralov, and Fedotova]{Altyntsev2020multiwave}
A.T. Altyntsev, S.V. Lesovoi, M.V. Globa, A.V. Gubin, A.A. Kochanov, V.V.
  Grechnev, E.F. Ivanov, V.S. Kobets, N.S. Meshalkina, A.A. Muratov, D.V.
  Prosovetsky, I.I. Myshyakov, A.M. Uralov, and A.Y. Fedotova.
\newblock Multiwave siberian radioheliograph.
\newblock \emph{Solar-Terrestrial Physics}, 6:\penalty0 30--40, 2020.

\bibitem[Grechnev et~al.(2014)Grechnev, Uralov, Chertok, Belov, Filippov,
  Slemzin, and Jackson]{Grechnev2014}
V.~V. Grechnev, A.~M. Uralov, I.~M. Chertok, A.~V. Belov, B.~P. Filippov, V.~A.
  Slemzin, and B.~V. Jackson.
\newblock A challenging solar eruptive event of 18 november 2003 and the causes
  of the 20 november geomagnetic superstorm. iv. unusual magnetic cloud and
  overall scenario.
\newblock \emph{Solar Physics}, 289\penalty0 (12):\penalty0 4653--4673, Dec
  2014.
\newblock ISSN 1573-093X.
\newblock \doi{10.1007/s11207-014-0596-5}.
\newblock URL \url{https://doi.org/10.1007/s11207-014-0596-5}.

\bibitem[Lesovoi et~al.(2017)Lesovoi, Altyntsev, Kochanov, Grechnev, Gubin,
  Zhdanov, Ivanov, Uralov, Kashapova, Kuznetsov, Meshalkina, and
  Sych]{Lesovoi2017siberian}
S.V. Lesovoi, A.T. Altyntsev, A.A. Kochanov, V.V. Grechnev, A.V. Gubin, D.A.
  Zhdanov, E.F. Ivanov, A.M. Uralov, L.K. Kashapova, A.A. Kuznetsov, N.S.
  Meshalkina, and R.A. Sych.
\newblock Siberian radioheliograph: first results.
\newblock \emph{Solar-Terrestrial Physics}, 3:\penalty0 3--18, 2017.

\bibitem[Picone et~al.(2002)Picone, Hedin, Drob, and Aikin]{picone}
J.~M. Picone, A.~E. Hedin, D.~P. Drob, and A.~C. Aikin.
\newblock Nrlmsise-00 empirical model of the atmosphere: Statistical
  comparisons and scientific issues.
\newblock \emph{Journal of Geophysical Research: Space Physics}, 107\penalty0
  (A12):\penalty0 SIA 15--1--SIA 15--16, 2002.
\newblock \doi{https://doi.org/10.1029/2002JA009430}.
\newblock URL
  \url{https://agupubs.onlinelibrary.wiley.com/doi/abs/10.1029/2002JA009430}.

\bibitem[Bowman et~al.(2008)Bowman, Tobiska, Marcos, Huang, Lin, and
  Burke]{Bowman2008}
Bruce Bowman, W.~Kent Tobiska, Frank Marcos, Cheryl Huang, Chin Lin, and
  William Burke.
\newblock \emph{A New Empirical Thermospheric Density Model JB2008 Using New
  Solar and Geomagnetic Indices}.
\newblock 2008.
\newblock \doi{10.2514/6.2008-6438}.
\newblock URL \url{https://arc.aiaa.org/doi/abs/10.2514/6.2008-6438}.

\bibitem[Bruinsma(2015)]{Bruinsma2015}
Sean Bruinsma.
\newblock The {{DTM-2013}} thermosphere model.
\newblock \emph{Journal of Space Weather and Space Climate}, 5:\penalty0 A1,
  2015.
\newblock ISSN 2115-7251.
\newblock \doi{10.1051/swsc/2015001}.

\bibitem[Zhang et~al.(2022)Zhang, Zhao, Feng, Liu, Xiang, Li, and
  Lu]{Zhang2022}
Wanting Zhang, Xinhua Zhao, Xueshang Feng, Cheng’ao Liu, Nanbin Xiang, Zheng
  Li, and Wei Lu.
\newblock Predicting the daily 10.7-cm solar radio flux using the long
  short-term memory method.
\newblock \emph{Universe}, 8\penalty0 (1), 2022.
\newblock ISSN 2218-1997.
\newblock \doi{10.3390/universe8010030}.
\newblock URL \url{https://www.mdpi.com/2218-1997/8/1/30}.

\bibitem[Stevenson et~al.(2022)Stevenson, Rodriguez-Fernandez, Minisci, and
  Camacho]{STEVENSON2022}
Emma Stevenson, Victor Rodriguez-Fernandez, Edmondo Minisci, and David Camacho.
\newblock A deep learning approach to solar radio flux forecasting.
\newblock \emph{Acta Astronautica}, 193:\penalty0 595--606, 2022.
\newblock ISSN 0094-5765.
\newblock \doi{https://doi.org/10.1016/j.actaastro.2021.08.004}.
\newblock URL
  \url{https://www.sciencedirect.com/science/article/pii/S009457652100415X}.

\bibitem[Hao et~al.(2024)Hao, Lu, Peng, Wang, Li, and Wei]{hao}
Yuhang Hao, Jianyong Lu, Guangshuai Peng, Ming Wang, Jingyuan Li, and Guanchun
  Wei.
\newblock F10.7 daily forecast using lstm combined with vmd method.
\newblock \emph{Space Weather}, 22\penalty0 (1):\penalty0 e2023SW003552, 2024.
\newblock \doi{https://doi.org/10.1029/2023SW003552}.
\newblock URL
  \url{https://agupubs.onlinelibrary.wiley.com/doi/abs/10.1029/2023SW003552}.
\newblock e2023SW003552 2023SW003552.

\bibitem[Egorov(2025)]{EGOROV2025455}
Yaroslav Egorov.
\newblock Forecasting f10.7 and f30 indices using the variational mode
  decomposition algorithm and the deep-learning time-series dense encoder
  model.
\newblock \emph{Acta Astronautica}, 234:\penalty0 455--461, 2025.
\newblock ISSN 0094-5765.
\newblock \doi{https://doi.org/10.1016/j.actaastro.2025.04.027}.
\newblock URL
  \url{https://www.sciencedirect.com/science/article/pii/S0094576525002279}.

\bibitem[Fedotova et~al.(2019)Fedotova, Altyntsev, Kochanov, Lesovoi, and
  Meshalkina]{Fedotova2019calibration}
A.Y. Fedotova, A.T. Altyntsev, A.A. Kochanov, S.V. Lesovoi, and N.S.
  Meshalkina.
\newblock Calibration of siberian radioheliograph images.
\newblock \emph{Solar-Terrestrial Physics}, 5:\penalty0 27--33, 2019.

\bibitem[Globa and Lesovoi(2021)]{Globa2021calibration}
M.V. Globa and S.V. Lesovoi.
\newblock Calibration of siberian radioheliograph antenna gains using
  redundancy.
\newblock \emph{Solar-Terrestrial Physics}, 7:\penalty0 98--103, 2021.

\bibitem[{H{\"o}gbom}(1974)]{Hogbom1974}
J.~A. {H{\"o}gbom}.
\newblock {Aperture Synthesis with a Non-Regular Distribution of Interferometer
  Baselines}.
\newblock \emph{Astronomy and Astrophysics Supplement}, 15:\penalty0 417, June
  1974.

\bibitem[Yu and Ma(2021)]{Yu2021}
Siwei Yu and Jianwei Ma.
\newblock Deep learning for geophysics: Current and future trends.
\newblock \emph{Reviews of Geophysics}, 59\penalty0 (3):\penalty0
  e2021RG000742, 2021.
\newblock \doi{https://doi.org/10.1029/2021RG000742}.
\newblock URL
  \url{https://agupubs.onlinelibrary.wiley.com/doi/abs/10.1029/2021RG000742}.
\newblock e2021RG000742 2021RG000742.

\bibitem[Asensio~Ramos et~al.(2023)Asensio~Ramos, Cheung, Chifu, and
  Gafeira]{Ramos2023}
Andr{\'e}s Asensio~Ramos, Mark C.~M. Cheung, Iulia Chifu, and Ricardo Gafeira.
\newblock Machine learning in solar physics.
\newblock \emph{Living Reviews in Solar Physics}, 20\penalty0 (1):\penalty0 4,
  Jul 2023.
\newblock ISSN 1614-4961.
\newblock \doi{10.1007/s41116-023-00038-x}.
\newblock URL \url{https://doi.org/10.1007/s41116-023-00038-x}.

\bibitem[Illarionov and Tlatov(2018)]{Illarionov2018}
Egor~A Illarionov and Andrey~G Tlatov.
\newblock Segmentation of coronal holes in solar disc images with a
  convolutional neural network.
\newblock \emph{Monthly Notices of the Royal Astronomical Society},
  481\penalty0 (4):\penalty0 5014--5021, 10 2018.
\newblock ISSN 0035-8711.
\newblock \doi{10.1093/mnras/sty2628}.
\newblock URL \url{https://doi.org/10.1093/mnras/sty2628}.

\bibitem[{Jarolim, R.} et~al.(2021){Jarolim, R.}, {Veronig, A. M.},
  {Hofmeister, S.}, {Heinemann, S. G.}, {Temmer, M.}, {Podladchikova, T.}, and
  {Dissauer, K.}]{Jarolim2021}
{Jarolim, R.}, {Veronig, A. M.}, {Hofmeister, S.}, {Heinemann, S. G.}, {Temmer,
  M.}, {Podladchikova, T.}, and {Dissauer, K.}
\newblock Multi-channel coronal hole detection with convolutional neural
  networks.
\newblock \emph{Astronomy and Astrophysics}, 652:\penalty0 A13, 2021.
\newblock \doi{10.1051/0004-6361/202140640}.
\newblock URL \url{https://doi.org/10.1051/0004-6361/202140640}.

\bibitem[Armstrong and Fletcher(2019)]{Armstrong2019}
John~A. Armstrong and Lyndsay Fletcher.
\newblock Fast solar image classification using deep learning and its
  importance for automation in solar physics.
\newblock \emph{Solar Physics}, 294\penalty0 (6):\penalty0 80, Jun 2019.
\newblock ISSN 1573-093X.
\newblock \doi{10.1007/s11207-019-1473-z}.
\newblock URL \url{https://doi.org/10.1007/s11207-019-1473-z}.

\bibitem[Siahkoohi et~al.(2019)Siahkoohi, Louboutin, and
  Herrmann]{Siahkoohi2019}
Ali Siahkoohi, Mathias Louboutin, and Felix~J. Herrmann.
\newblock The importance of transfer learning in seismic modeling and imaging.
\newblock \emph{Geophysics}, 84\penalty0 (6):\penalty0 A47--A52, 10 2019.
\newblock ISSN 0016-8033.
\newblock \doi{10.1190/geo2019-0056.1}.
\newblock URL \url{https://doi.org/10.1190/geo2019-0056.1}.

\bibitem[Hosna et~al.(2022)Hosna, Merry, Gyalmo, Alom, Aung, and
  Azim]{Hosna2022}
Asmaul Hosna, Ethel Merry, Jigmey Gyalmo, Zulfikar Alom, Zeyar Aung, and
  Mohammad~Abdul Azim.
\newblock Transfer learning: a friendly introduction.
\newblock \emph{Journal of Big Data}, 9\penalty0 (1):\penalty0 102, Oct 2022.
\newblock ISSN 2196-1115.
\newblock \doi{10.1186/s40537-022-00652-w}.
\newblock URL \url{https://doi.org/10.1186/s40537-022-00652-w}.

\bibitem[Radford et~al.(2021)Radford, Kim, Hallacy, Ramesh, Goh, Agarwal,
  Sastry, Askell, Mishkin, Clark, Krueger, and Sutskever]{clip}
Alec Radford, Jong~Wook Kim, Chris Hallacy, Aditya Ramesh, Gabriel Goh,
  Sandhini Agarwal, Girish Sastry, Amanda Askell, Pamela Mishkin, Jack Clark,
  Gretchen Krueger, and Ilya Sutskever.
\newblock Learning transferable visual models from natural language
  supervision.
\newblock \emph{CoRR}, abs/2103.00020, 2021.
\newblock URL \url{https://arxiv.org/abs/2103.00020}.

\bibitem[Tan and Le(2020)]{tan2020efficientnetrethinkingmodelscaling}
Mingxing Tan and Quoc~V. Le.
\newblock Efficientnet: Rethinking model scaling for convolutional neural
  networks, 2020.
\newblock URL \url{https://arxiv.org/abs/1905.11946}.

\bibitem[Dorogush et~al.(2018)Dorogush, Ershov, and Gulin]{catboost}
Anna~Veronika Dorogush, Vasily Ershov, and Andrey Gulin.
\newblock Catboost: gradient boosting with categorical features support.
\newblock \emph{CoRR}, abs/1810.11363, 2018.
\newblock URL \url{http://arxiv.org/abs/1810.11363}.

\bibitem[Tan and Le(2019)]{effnet}
Mingxing Tan and Quoc~V. Le.
\newblock Efficientnet: Rethinking model scaling for convolutional neural
  networks.
\newblock \emph{CoRR}, abs/1905.11946, 2019.
\newblock URL \url{http://arxiv.org/abs/1905.11946}.

\end{thebibliography}

\end{document}